# Zero valley splitting at zero magnetic field for strained Si/SiGe quantum wells grown on tilted substrates

Paul von Allmen and Seungwon Lee

*Jet Propulsion Laboratory, California Institute of Technology, Pasadena, CA 91109*

(Dated: June 14, 2006)

**Abstract**

The electronic structure for a strained Si/SiGe quantum well grown on a tilted substrate with periodic steps is calculated using a parameterized tight-binding method. For a zero tilt angle the energy difference between the two lowest minima of the conduction band at the center of the Brillouin zone defines a non-zero valley splitting. At finite tilt angles, the two lowest conduction band minima shift to $k_0$ and $-k_0$ in the Brillouin zone and have equal energy. The valley splitting for quantum wells grown on a tilted substrate is therefore equal to zero, which is a direct consequence of the periodicity of the steps at the interfaces between the quantum well and the buffer materials.

When a strained silicon quantum well (QW) is grown on top of a relaxed SiGe buffer, the Z valley bands (direction perpendicular to the surface) are split from the X and Y valleys to lower energy [1]. The confinement of the electrons in the QW induces an additional splitting of the two Z valley states. This splitting is termed valley splitting (VS) and has been predicted, computed and measured many times over the past decades for a number of silicon structures [2-4]. Understanding the origin and size of VS has become more important over the past years for the role that VS plays in defining a qubit in silicon quantum dots for quantum computing applications [5]. In a silicon-based quantum computer, a large VS is required to unambiguously define the qubit Hilbert space using the two spin states of the lowest Z valley level.

Calculations within the effective mass approximation have shown that the VS is strongly suppressed if the QW is grown on a tilted substrate compared to the QW with no tilt [2]. A first order perturbation calculation shows that the VS is zero at zero magnetic field if the steps resulting from the growth on a tilted substrate are periodically repeated [6]. However, a more involved variational calculation that includes charge density oscillations predicts that the residual VS at B=0 is non-zero [6]. The intuitive explanation is that the destructive interference between the contributions to the VS from the periodically repeated steps is incomplete because the charge density oscillations due to the steps are incommensurate with the crystal-induced oscillations. The effective mass calculation uses ad hoc parameters to describe the interface potential and the charge density oscillations ("washboard" potential). The present paper reports VS calculations for a silicon QW on a tilted substrate obtained with a parameterized tight-binding method, where no ad hoc parameters are added to describe the interface and the charge density oscillations. The main result is that the VS is zero at B=0 despite the charge density oscillations, at variance with Ref. [6].

The electronic structure for the QW is calculated using a parameterized tight-binding method that has been widely applied to the modeling of semiconductor bulk materials and nanostructures [3, 6, 8]. Bulk materials are described with a one-particle Schrödinger equation, and the wavefunctions are expanded on a basis set of orthogonalized atomic orbitals (Löwdin orbitals). The matrix elements of the Hamiltonian are restricted to nearest-neighbor interactions and tuned to fit a set of material parameters such as the energy band edges, the effective masses and the deformation potentials [7]. The parameters for silicon are taken from reference [7], and include the effects of strain in the off-diagonal elements of the Hamiltonian matrix. The confinement in the freestanding silicon QW is obtained by passivating the surface dangling bonds along the $sp^3$ bond directions [9]. The biaxial strain in the silicon QW due to the lattice mismatch between the Si and $Si_{1-x}Ge_x$ materials is taken as uniform with strain tensor values $\varepsilon_{xx} = \varepsilon_{yy} = 0.01253$ and $\varepsilon_{zz} = -0.01029$, which correspond to a Ge concentration $x = 0.3$. The Hamiltonian is diagonalized using an iterative Lanczos algorithm [10].

A silicon QW grown on top of a tilted substrate displays steps resulting from the misorientation between the crystal axis $z$ [001] and the electronic confinement direction $z'$ (Fig. 1). Such steps have been observed experimentally and found to form irregular patterns with however a good measure of alignment along preferential directions [11]. As an initial stage in the atomic-level modeling of these interfaces, we examine a structure with mono-atomic steps aligned along the direction [100], with height $h = a/4$, where $a = 0.543$ nm is the silicon lattice constant. The width of the steps $L_s$ is related to the tilt angle of the substrate, $\theta$ by $\tan\theta = h/L_s$. A key feature of this structure is that the steps are repeated periodically in direction $x'$, parallel to the QW layer. Since the step height is $a/4$, the atomic structure is repeated periodically after four steps. Periodic boundary conditions can therefore be applied to the tight-binding Hamiltonian with periodicity along $x'$ $P = \sqrt{16L_s^2 + a^2}$. Since the structure is also periodic along direction y [010], the eigenvalues of the Schrödinger equation are described by a two-dimensional band structure with a Brillouin zone defined by $\{(k_{x'}, k_y) | -\pi/P \leq k_{x'} \leq \pi/P, -\pi/a \leq k_y \leq \pi/a\}$.

The band structure for a QW without tilt is first computed to serve as a reference point for further calculations. Figure 2 shows the band structure along $x$ (parallel to the QW layer) for a $L_z = 5.43$ nm wide freestanding strained Si QW with passivated dangling bonds at the surface and with growth direction [001] parallel to the confinement direction (no tilt). The two lowest minima at $k_x = 0$ correspond to the two Z valley states, the energy splitting of which (inset of Fig. 2) oscillates with QW thickness as reported previously both within the tight binding and the effective mass approximation methods [3, 6]. The minima at $k_x \cong \pm 1/a$ and energies about 130 meV higher than the Z-valley minima correspond to the states originating from X valleys.

Figure 3 shows the band structure along $x'$ (parallel to the tilted QW layer) for several tilt angles. Two minima with equal energies corresponding to the Z valley states are located at $k_{x'} = \pm k_{x'\min}(\theta)$, where $k_{x'\min}(\theta)$ increases monotonously with the tilt angle. The conduction band minima occur at $k_{x'} \neq 0$ because the bulk silicon conduction band minima are at $k \neq 0$ and because the confinement direction $z'$ is at an angle $\theta$ from the crystal z [001] direction. This result can be interpreted more explicitly in terms of simple arguments using the effective mass approximation. The eigenfunctions for the lowest Z-valley confined states for a QW of width $W$ and with infinite potential barriers can be approximated by the following expression

$$\psi^{\pm}(x',z') = \cos\left(\frac{\pi}{W}z'\right)e^{\pm ik_0\cos\theta z'}e^{ik_{x'}x'}.$$

The phase factor in the $z'$ direction is $k_0\cos\theta z'$ because the bulk conduction band minima in the rotated coordinate frame $(x'y'z')$ are at $k = \mp k_0\sin\theta\hat{e}_{x'} \pm k_0\cos\theta\hat{e}_{z'}$, where $k_0 \cong 0.15\,(2\pi/a)$ is the position of the conduction band minimum in the bulk silicon folded Brillouin zone. Introducing these wavefunctions into the effective mass equation for parabolic conduction bands with diagonal components of the effective mass tensor in the rotated frame, $m'_l$ and $m'_t$, yields the following expression for the eigenenergies:

$$E_c^{\pm}(k_{x'}) = E_{c,\min} + \frac{\hbar^2(k_{x'} \pm k_0\sin\theta)^2}{m'_t} + \frac{\hbar^2\left(\frac{\pi}{W}\right)^2}{m'_l} + O(\sin 2\theta).$$

The off-diagonal terms of the effective mass tensor, which are non-zero in the rotated frame and explicit contributions from the steps are neglected. The energy minima are obtained at $k^{\pm}_{x'\min} = \mp k_0\sin\theta$, in excellent agreement with the numerical results in Figure 3. The fact that the two minima have the same energy values indicate that the valley splitting is zero for a QW grown on a tilted substrate with periodic steps. This conclusion confirms published results using first order perturbation within the effective mass approximation but disagrees with reported variational calculations that conjecture that charge density oscillations along $x'$ with periodicity $L_s$ result in a non-zero VS at $B=0$ [6].

Figure 4 shows the electron probability density at the band minimum integrated over *y* and *z* as a function of *x*. An approximately 0.1% modulation is superposed over the atomic oscillations. This result confirms the hypothesis in Ref. [6] of a charge density oscillation contribution to the effective mass equation. However, our band calculation suggests that the charge oscillation does not yield a non-zero valley splitting, in contrast to the variational calculation result.

Figure 5 shows the *z*-variation of the electron probability density at the band minima integrated over *y* and for four different positions along x, located at the middle of the four steps. The maximum amplitude of the wavefunctions shifts with the center of the QW, showing that the confinement follows the tilted QW. This result supports the hypothesis made in the simple effective analysis above, where the maximum amplitude of the envelope function is taken at constant z'=0.

It should be noted that any perturbation of the periodicity of the interface steps will break the translational invariance along *x'* and will result in a finite VS. Examples of perturbations are a magnetic field, a lateral confinement electrostatically induced with surface gates, and fluctuations is the step geometry. The finite VS observed experimentally in Ref. [4] is therefore not in contradiction with the results reported here.

The band structure for a strained silicon QW grown on an unstrained SiGe substrate with periodic steps was calculated with an atomic-level model. The lowest two minima in the conduction band are located off-center in the Brillouin zone and have equal energy values. The VS is therefore exactly zero at zero magnetic field as a direct consequence of the periodicity of the steps at the interface.

**Acknowledgment**

We are grateful for fruitful discussions with Mark Friesen, Mark Eriksson, and Fabiano Oyafuso. This work was performed at the Jet Propulsion Laboratory, California Institute of Technology under a contract with the National Aeronautics and Space Administration. Funding was provided under a grant from NSA.

**References**

[1] C. Herring and E. Vogt, Phys. Rev. **101**, 944 (1956).

[2] T. Ando, Phys. Rev. B 19, 3089 (1979) and references therein.

[3] T.B. Boykin, G. Klimeck, M.A. Eriksson, M. Friesen, S.N. Coppersmith, P. von Allmen, F. Oyafuso, and S. Lee, Appl. Phys. Lett. 84, 115 (2004); T.B. Boykin, G. Klimeck, M. Friesen, S.N. Coppersmith, P. von Allmen, F. Oyafuso, and S. Lee, Phys. Rev. B 70, 165325 (2004).

[4] S. Goswami, M. Friesen, J.L. Truitt, Ch. Tahan, L.J. Klein, J.O. Chu, P.M. Mooney, D.W. van der Weide, S.N. Coppersmith, R. Joynt, and M. Eriksson, cond-mat/0408389v4 (2006) and references therein.

[5] M. Friesen, P. Rugheimer, D. E. Savage, M. G. Lagally, D. W. van der Weide, R. Joynt, and M. Eriksson, Phys. Rev. B 67, 121301 (2003).

[6] M. Friesen, M.A. Eriksson, and S.N. Coppersmith, cond-mat/0602194v2 (2006).

[7] J.-M. Jancu, R. Scholtz, F. Beltram, and F. Bassani, Phys. Rev. B **57**, 6493 (1998).

[8] S. Lee and P. von Allmen, Appl. Phys. Lett. 88, 022107 (2006).

[9] S. Lee, F. Oyafuso, P. von Allmen, and G. Klimeck, Phys. Rev. B 69, 045316 (2004).

[10] *"Iterative Methods for Sparse Linear Systems"* Y. Saad, PWS Publishing Company, Boston 1996.

[11] B.S. Swartzentruber, Y.-W. Mo, M.B. Webb, and M.G. Lagally, J. Vac. Sci. Technol. A7, 2901 (1989).

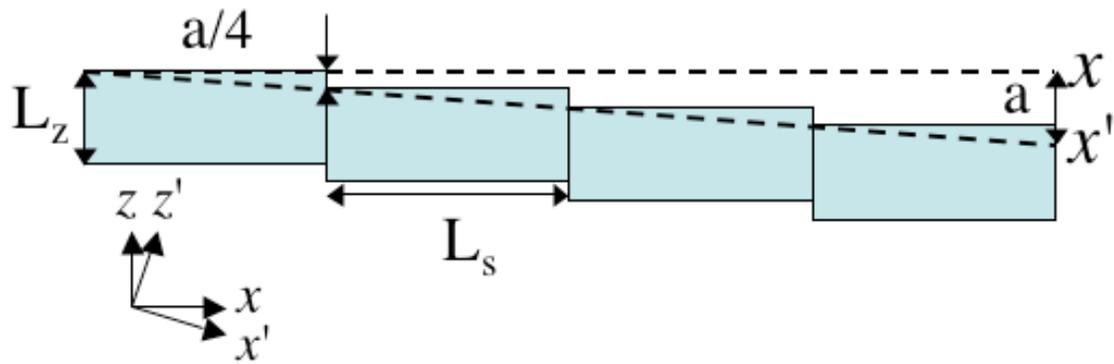

**Figure 1:** Geometry of a quantum well grown on a tilted substrate. The crystal symmetry directions are along *x* and *z*. The QW confinement direction is along *z'* and *x'* is in the plane of the QW. The step height is one atomic layer (*a*/4), and the atomic structure is periodic after four steps with a displacement along *z* of one full unit cell length (*a*).

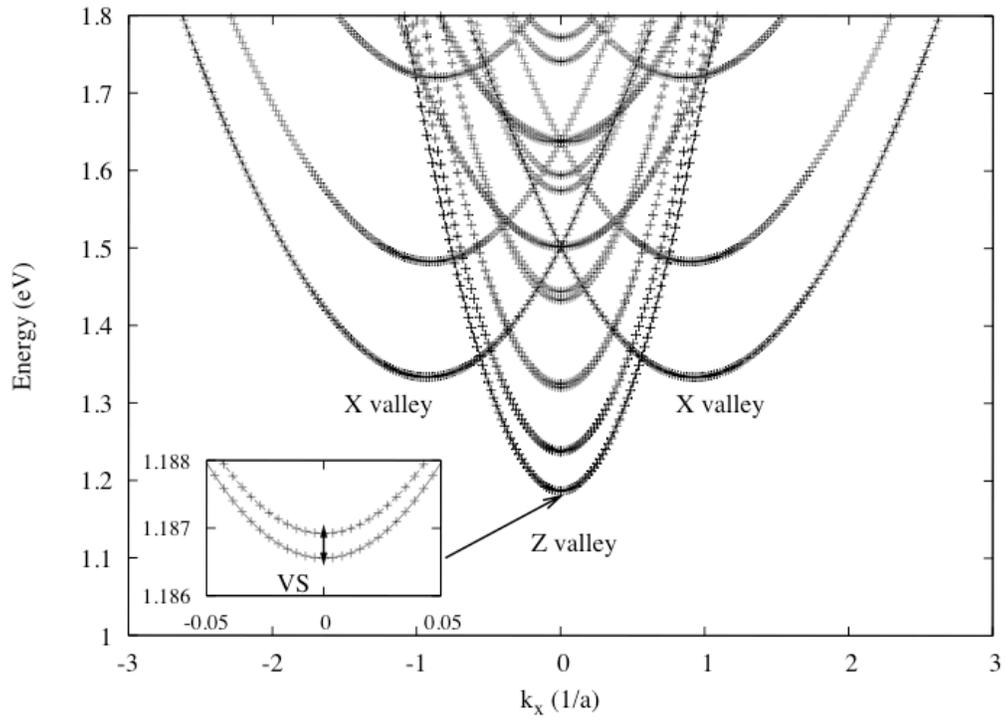

**Figure 2:** Band structure of a 5.43 nm wide strained silicon QW with no tilt. The inset shows the valley splitting at the band minimum $k_x = 0$.

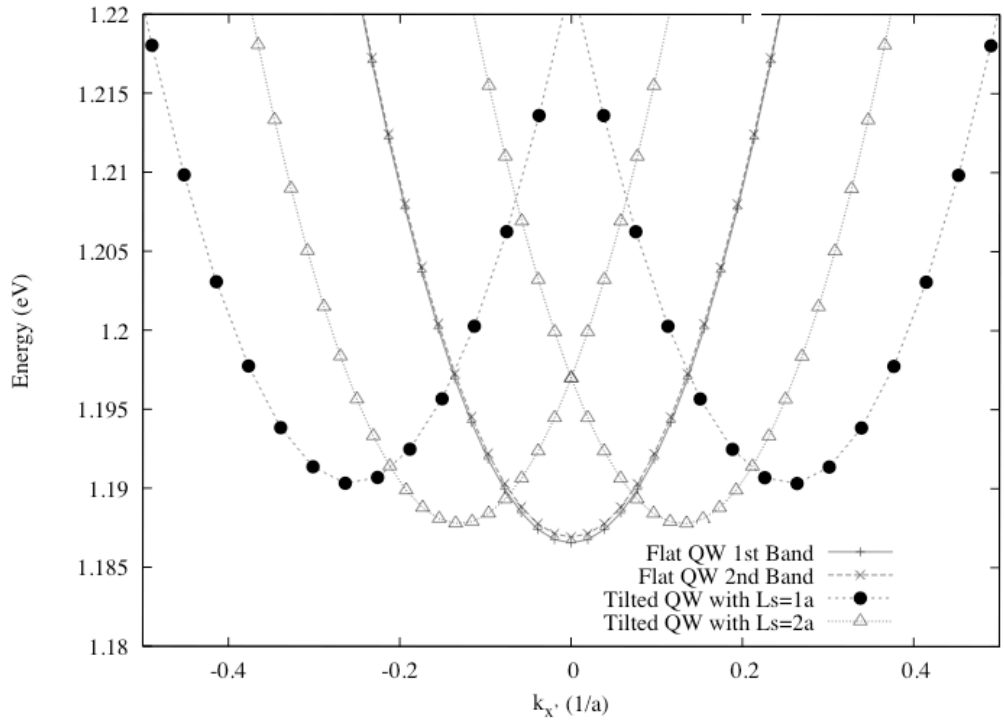

**Figure 3:** Band structure for a tilted 5.43 nm wide strained silicon QW for several tilt angles (tilt angle is smaller when $L_s$ is larger). For finite tilt angles the band minima are at $k_{x'} \neq 0$ (see text).

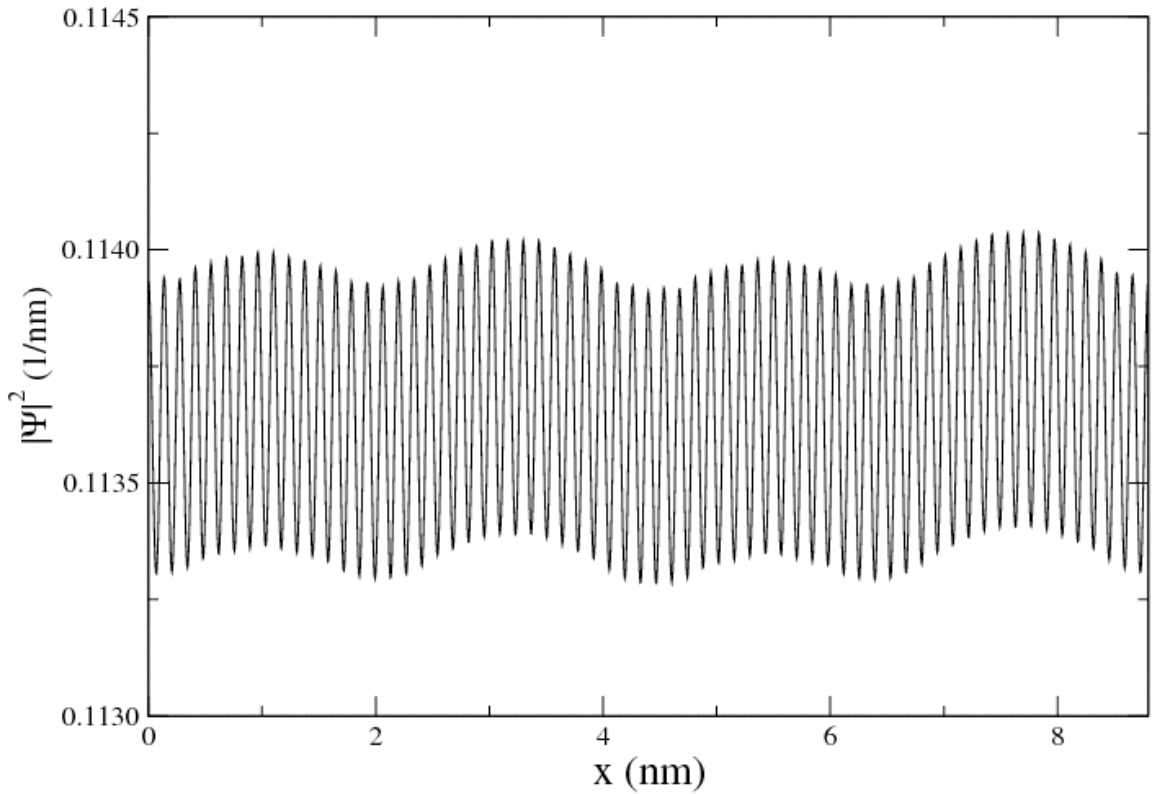

**Figure 4:** Electron probability distribution at the band minimum integrated over *y* and *z*. The QW is 5.43 nm wide and grown on a tilted substrate with step size *L$_s$=4a*. The probability distribution is obtained from the tight-binding wave function, which is a linear combination of atomic orbitals. In order to display the probability distribution as a continuously varying function, the atomic orbitals are approximated with Gaussian-type orbitals.

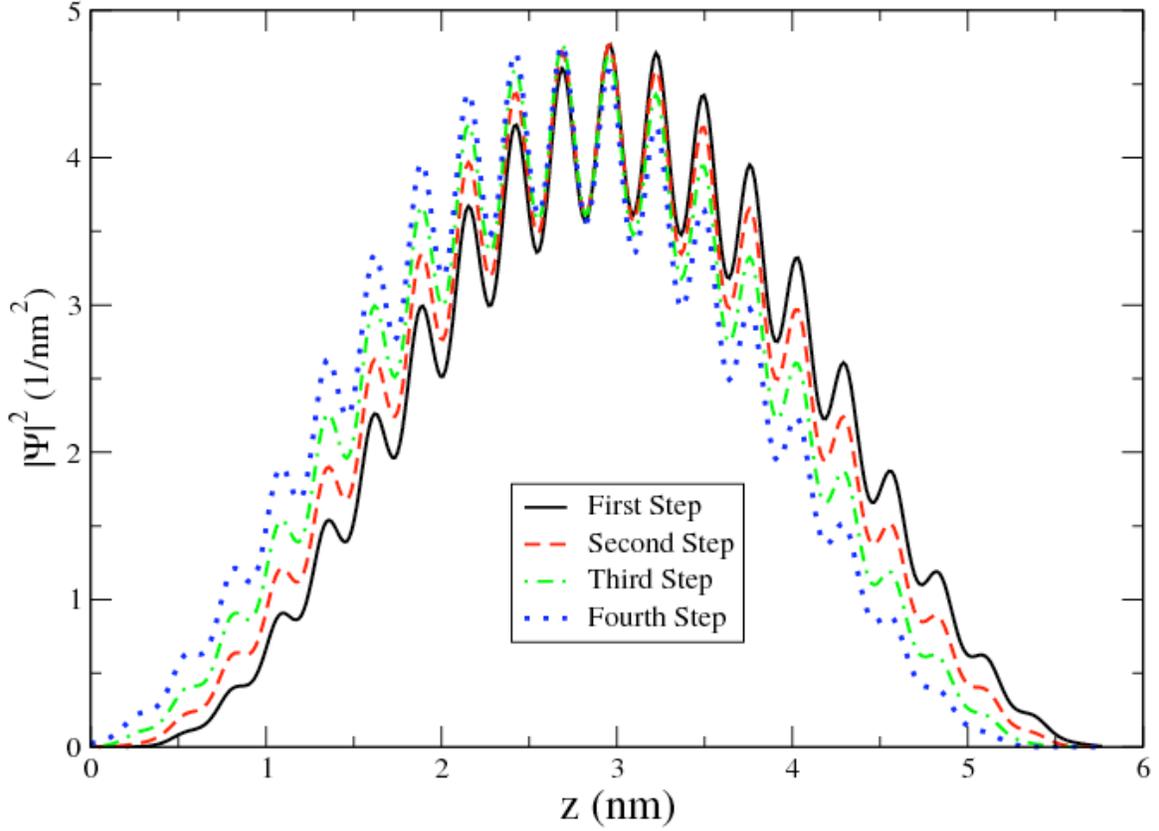

**Figure 5:** Electron probability distribution at the band minimum integrated over *y* and at different locations along *x*. The QW is 5.43 nm wide and grown on a tilted substrate with step size $L_s=4a$. The continuously varying distribution is obtained by approximating the tight-binding basis orbitals with Gaussian-type orbitals.